\documentstyle[11pt,paspconf,epsf]{article}

\begin{document}

\vspace*{-25mm}
\noindent
To appear in {\em The Low Surface Brightness Universe}, the proceedings 
of I.A.U. Colloquium No.~171, Cardiff, 6th--10th July, 1998, 
eds. J.~I.~Davies, C.~D.~Impey \& S.~Phillipps, publ. Astronomical 
Society of the Pacific, in press.         
\\[11mm]

\title{Low Surface Brightness Dwarf Galaxies in the Bristol -- 
    Anglo-Australian Observatory Virgo Cluster Survey} 

\author{J. B. Jones, S. Phillipps, J. M. Schwartzenberg}
\affil{Astrophysics Group,  Department of Physics, 
    University of Bristol, Tyndall Avenue, Bristol, BS8 1TL, 
    United Kingdom.}
\author{Q. A. Parker}
\affil{Anglo-Australian Observatory, Siding Spring, Coonabarabran,
    New South Wales 2357, Australia.}

\begin{abstract}
    We describe a new, deep photographic survey of the Virgo 
Cluster which uses multiple exposures on Tech Pan film 
with the United Kingdom Schmidt 
Telescope to probe the dwarf population to fainter surface brightness 
limits than previous surveys. We have identified galaxies having 
sizes ($\geq 3$~arcsec scale length) and surface brightnesses 
($\leq 24.5$~R~mag~arcsec$^{-2}$) characteristic of those expected for 
dwarf spheroidal galaxies in the cluster. 
The survey is providing substantial samples of extremely low luminosity 
galaxies outside the environment of the Local Group and nearby groups 
for the first time. 
An initial study of two small areas has found dwarf spheroidal 
candidates in large numbers (500 deg$^{-2}$) which 
indicate a steep, continuously rising luminosity function at 
low luminosities. 
\end{abstract}


\section{The Virgo Cluster}

   The Virgo Cluster, alongside its counterpart in Fornax, 
is the nearest sizeable galaxy cluster. It is close enough for 
detailed morphological studies to be possible even for low
luminosity dwarf galaxies. 
It is an irregular, poor cluster of Bautz-Morgan type III 
(Abell 1975) and Abell richness class 0. As such, it 
allows detailed studies of the dwarf population 
in an environment substantially different from the Local Group 
and other nearby groups. 

   A seminal study was carried out by Binggeli, Sandage and 
Tammann, who generated the Virgo Cluster Catalog consisting 
of 1277 galaxies classified as certain members and a further 
574 possible members over an area of 140~deg$^2$ (Binggeli et al. 1985). 
Membership was assigned by visual inspection, 
essentially based on the larger angular sizes of the 
cluster galaxies compared with the background population. 
Their dwarfs conformed to a moderately steep luminosity function 
(Sandage et al. 1985, Binggeli et al. 1988). 
Various detailed studies of cluster members have been performed 
subsequently, including the dwarf population (e.g. Ferguson \& 
Sandage 1989, Binggeli \& Cameron 1993, Durrell 1997, 
Young \& Currie 1998). 

   Of particular note, Impey, Bothun \& Malin (1988) performed 
a survey for large angular size low surface brightness galaxies 
in a single Schmidt field centred on the M87 cluster core, 
using a photographic stacking technique. They identified 137 
galaxies having central surface brightnesses in the range 23 to 
26~B~mag~arcsec$^{-2}$, of which 27 were new detections.

\section{The Bristol -- AAO Virgo Cluster Survey}

    The properties of the faint end of the galaxy luminosity function 
are poorly constrained outside the Local Group, both in terms 
of the numbers and characteristics of the galaxies. Some studies 
have found evidence for very steep luminosity functions in both 
cluster (e.g. De Propris et al. 1995, Smith, Driver \& 
Phillipps 1997, Trentham 1997, 1998a) and field (Loveday 1997, 
Morgan, Smith \& Phillipps 1998) environments. At the very lowest 
luminosities, dwarf spheroidal 
galaxies have been identified in nearby groups (e.g. Caldwell 
et al. 1998), extending the number of known dSphs and allowing 
a comparison of their properties with Local Group members. 
The importance of very low luminosity galaxies in understanding 
galaxy formation (e.g. Frenk et al. 1996, Kauffmann, 
Nusser \& Steinmetz 1997) and evolution (Gallagher \& Wyse 1994, 
Caldwell et al. 1998, Trentham 1998b) demands that progress 
is made in identifying and studying extremely low luminosity dwarfs 
in new environments (see Phillipps et al. 1998a, and references 
therein). 

   In order to extend surveys of galaxies to fainter surface 
brightnesses than previous surveys 
and over a full ten degree square region of the cluster, the 
Bristol--Anglo-Australian Observatory survey is using multiple 
exposures with the United Kingdom Schmidt Telescope (UKST) on Kodak 
Tech Pan emulsion through an R band filter 
(Schwartzenberg, Phillipps \& Parker 1995, Schwartzenberg \& 
Phillipps 1995, 1997, Schwart\-zenberg 1996, Phillipps et al. 1998b, 
Jones et al. 1998). 
Six individual exposures of each Virgo field of 1 -- $1\frac{1}{2}$ 
hour duration are digitally stacked to give a total integration 
time of 7 hours. 
The Tech Pan emulsion combines a high efficiency (approaching 
10\%, Parker et al. 1998, Phillipps \& Parker 1993) and fine grains 
(providing a high imaging resolution of 5~microns and a high uniformity). 

\begin{figure}[!htb]
\vspace*{-30mm}
\plotone{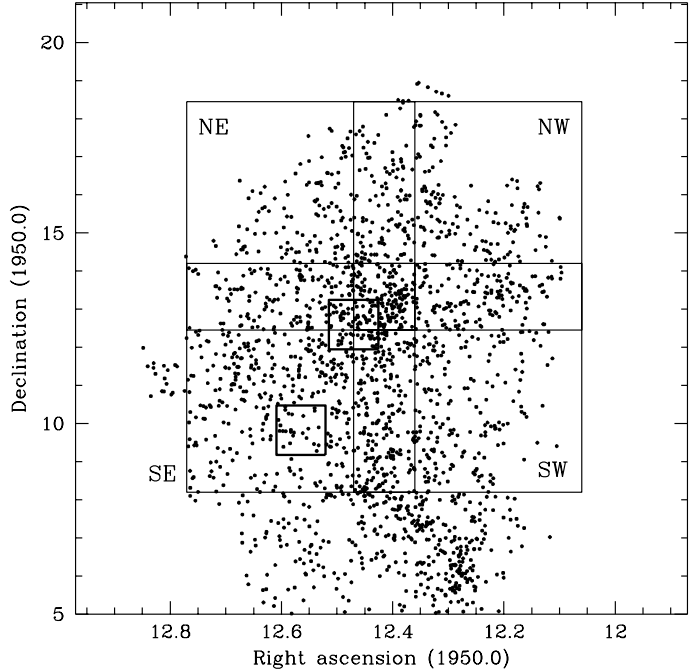}
\vspace*{-7mm}
\caption{
The four UKST fields and the Virgo Cluster. The boundaries 
of the four Schmidt fields are drawn superimposed on the 
distribution of cluster members given in the Virgo Cluster Catalog 
of Binggeli, Sandage \& 
Tammann (1985). The full 6~deg $\times$ 6~deg extent of the fields 
are shown; in practice the scanned regions are slightly smaller. 
The grid of fields is centred on the M87 core of the cluster. 
The M49 condensation lies at the southern boundary of the 
surveyed region. The small squares show the fields used in the 
initial survey of Section~5. 
}\label{fig-1}
\vspace*{10mm}
\end{figure}

   The four UKST fields of the Virgo survey area are shown in 
Figure~\ref{fig-1}. Field 
coordinates have been selected so as to cover a 100~deg$^2$ 
region centred on the M87 cluster core (``Cluster~A'' of 
Binggeli, Sandage \& Tammann 1987). The data therefore survey 
both the region 
of high density in the core and the lower density regions at 
the periphery. The survey area extends south as far as the M49 
cluster (``Cluster B'') and includes the ``M Cloud'' west of 
the M87 cluster. The central area in the grid has been imaged 
on all 24 films, in effect providing a further 1~mag depth gain in 
this region and offering the opportunity of a deeper survey in a 
restricted area. 
Table~\ref{tab-1} presents details of the four fields.

\begin{table}
\caption{The four quadrants of the surveyed area.} \label{tab-1}
\begin{center}
\begin{tabular}{lcccl}
\hspace*{3mm} Field & \multicolumn{2}{c}{Central coordinates} & 
No. exposures & Total integ. \\
      & R.A. (1950.0) & Dec. (1950.0) & used &  \hspace{4.5mm} time \\[1mm]
\tableline \\[-2.5mm]
Northwest & \hspace*{1.4mm}12$^{\rm h}$ 16$^{\rm m}$ & ~+15.4$^\circ$ & 6 & 
   \hspace{4mm} 6.3 hr \\[1mm]
Southwest & 12~~ 16~ & +11.2 & 6 & \hspace{4mm} 6.0 \\[1mm]
Northeast & 12~~ 34~ & +15.4 & 6 & \hspace{4mm} 6.3 $\;^{1}$ \\[1mm]
Southeast & 12~~ 34~ & +11.2 & 6 & \hspace{4mm} 7.1 \\[1mm]
\end{tabular}
\mbox{ }\\[3mm]
$\;^{1}$ Estimated (one exposure still awaited). 
\end{center}
\end{table}

    The photographic study compares very favourably with any CCD surveys 
currently feasible. The very large solid 
angles of the Schmidt fields, and the very long integration times, 
overcome the modest telescope aperture (1.2m) and the low quantum 
efficiency of the emulsion compared with CCDs. It has been 
possible to perform a very deep survey of a 10~deg $\times$ 10~deg region 
in the central part of the Virgo Cluster in the equivalent of 
only four nights' observing on the Schmidt. An equivalent survey 
to the same depth using CCDs might reasonably take several weeks, 
dependent on the details of detector and telescope. To date, the 
photographic observations are $96\%$ complete.

\section{Data Reduction and Analysis}

    The photographic data are digitised using the SuperCOSMOS 
plate measuring machine at the Royal Observatory Edinburgh. 
Only  the best quality films are used for stacking: adding poor 
quality films does not provide a useful gain in depth 
(Bland-Hawthorn, Shopbell \& Malin, 1993). 
With 10~micron wide pixels, corresponding to 0.67~arcsec on the sky, 
data extents are very large, causing 
particular data processing difficulties. 

    The individual exposures are sky-subtracted using a pixel-by-pixel 
subtraction of a median-filtered version of the data. The use of a 
3~arcmin square spatial filter maximises the ability to detect small, 
faint images but limits the survey, at least at the present time, 
to small galaxy images -- for example, the larger galaxies in the 
survey of Impey, Bothun and Malin (1988) would be partially 
removed. 
All six films of a given part of sky have their intensities normalised 
to the same scale, thus correcting for factors such as differences in 
exposure time, atmospheric transmission or the details of the 
development process. 
Coaddition of the six exposures is accomplished using median 
stacking, which efficiently overcomes problems such as film defects, 
dust particle images or satellite trails that affect single 
films. Image detection is performed with a connected-pixel 
algorithm using a detection threshold 
$\mu_{lim} = 25.5$~R~mag~arcsec$^{-2}$ and a minimum image 
area above this isophote of $A_{lim} = 11\:\mbox{arcsec}^2$. 
These parameters ensure that each detection will have a signal-to-noise 
ratio of least 10 and magnitudes $R \leq 22$~mag.

\section{Identifying Virgo Dwarf Spheroidals}

    Whilst the detection of galaxies in the direction of the 
Virgo Cluster may not present a particular challenge in itself, 
the identification of cluster members is significantly more 
difficult. Nearby clusters cover large angular scales and 
consequently the cluster population is swamped by the numerically 
dominant background population (see, for example, the review 
by Trentham 1998b). In the absence of redshift information 
(for example because of the practicalities of performing spectroscopy for 
very large samples of galaxies extending over wide areas, or because 
of the difficulties of obtaining spectra for low surface brightness
galaxies), membership for the dwarf population must often be assigned on 
morphological grounds. 
Binggeli, Sandage \& Tammann (1985) were able to assign membership for 
dwarf galaxies in the cluster on the basis of the galaxies' visual 
appearance: the dwarfs are characterised by their low surface 
brightnesses for their sizes. 

In the present survey, even though 
the galaxies are more extreme than those of the Virgo Cluster 
Catalog, it is still possible to isolate samples of cluster 
galaxies likely to suffer only a small degree of 
contamination by the background population. A comparison of Virgo 
photographic data with deep CCD data from a South Galactic Pole 
field shows that the background contamination is as small as several 
percent for galaxies having central surface brightnesses in the 
range $\mu_0 = 22.0$ to 24.5~R~mag~arcsec$^{-2}$ and exponential 
scale lengths $a \ge 3$~arcsec, even when allowance is made for 
differences in resolution between the two data sets. These issues 
are discussed in greater detail by Schwartzenberg (1996) 
and by Jones et al. (1998).

    The galaxies selected have properties comparable to those that 
Local Group dwarf spheroidals would have if they were at the 
distance of the Virgo Cluster. They have exponential scale lengths 
$h\geq 260$~pc and absolute magnitudes $M_{\rm R} = -11$ to $-16$ 
(for an assumed distance modulus of 31.3~mag, equivalent to 
$H_0 = 70$~km$\:\mbox{s}^{-1}\:\mbox{Mpc}^{-1}$). 
Local Group dwarf spheroidals have sizes between $h=90$ and 400~pc and 
absolute magnitudes $M_{\rm R} = -9$ to $-14$. The Virgo objects 
generally have fainter surface brightnesses than the galaxies of 
Binggeli, Sandage and Tammann (1985) and are smaller than those 
of Impey, Bothun and Malin (1988).

\section{The Initial Survey}

     A preliminary study of two subfields in the Southeast quadrant 
of the full survey area 
has been carried out. The SuperCOSMOS scans of the quadrant 
were provided as nine sections, each $7680 \times 7680$~pixels 
in size, two of which were used for the initial survey 
(Schwartzenberg et al., 1995). 
Excluding their edges, both fields are 1.3~deg square, providing 
a total area of 3.2~deg$^2$. The fields, listed in Table~\ref{tab-2}, 
were selected to sample the core of the cluster and a region of lower 
density 3.1~deg to the southeast. The core field included M87, 
although due to the raised background light levels, the region 
immediately around M87 was excluded from the study. 
The data were further subdivided into $2280 \times 2280$~pixel 
subregions for data processing because of computer hardware 
limitations. Fuller details are given by Schwartzenberg (1996) 
and Phillipps et al. (1998b).

\begin{table}
\label{tab-2}
\caption{Details of the fields studied in the initial survey.} 
\begin{center}
\begin{tabular}{llc}
\hspace*{3mm} Property & Core field & Outer field \\[1mm]
\tableline \\[-2.5mm]
Field area & ~1.58 deg$^{2}$ & ~1.61 deg$^{2}$ \\[1mm]
Field centre: & & \\[1mm]
\hspace*{4mm} R.A.(1950)&  $12^{\rm h} \; 28.2^{\rm m}$ & 
   ~$12^{\rm h} \; 33.9^{\rm m}$   \\[1mm]
\hspace*{4mm} Dec.(1950) & +12$^{\circ}\; 36'$ & +09$^{\circ}\; 49'$ \\[1mm]
\end{tabular}
\end{center}
\end{table}

\begin{figure}[!htb]
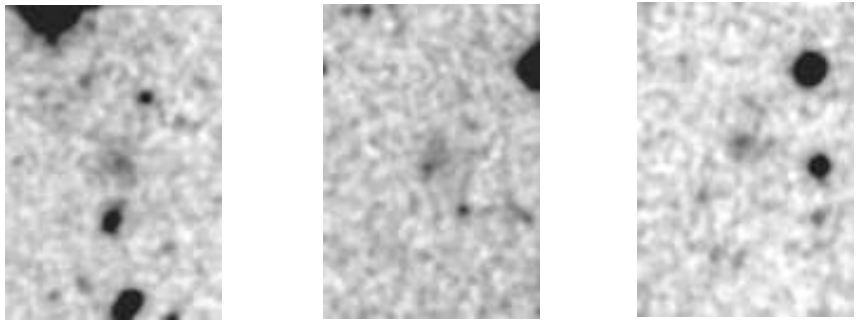

\vspace*{-50mm}
\plotfiddle{fig2a.eps}{90mm}{0}{15}{15}{-170}{0}
\plotfiddle{fig2b.eps}{0mm}{0}{15}{15}{-50}{25}
\plotfiddle{fig2c.eps}{0mm}{0}{15}{16}{70}{50}
\vspace*{-8mm}
\caption{
Examples of candidate Virgo Cluster dwarf spheroidal galaxies 
from the initial survey of 3.2~deg$^{2}$. The images have been 
produced by coadding six UKST exposures on Tech Pan film. 
None of the galaxies appears in either the Virgo Cluster 
Catalog or the sample of Impey, Bothun \& Malin (1988). 
Each frame is 1.0~arcmin in width. North is at the top. 
}\label{fig-2}
\vspace*{10mm}
\end{figure}

     A total of 56~000 images were detected over the two fields. 
From these a subsample of galaxies was selected having central surface 
brightnesses in the range $\mu_0 = 22.0$ to 24.5~R~mag~arcsec$^{-2}$ 
and scale lengths $a \ge 3$~arcsec after fitting exponential 
light profiles to the data. 
Through a detailed comparison of the numbers of Virgo galaxies with 
the South Galactic Pole field population at each 
point in the magnitude -- surface brightness plane, it is 
possible to remove the background contamination. 
The overall contamination of the sample is expected to be 
8~per cent. After this background subtraction, the sample contains 
1570 galaxies across the two fields. 
A luminosity function can be constructed 
by binning the background-subtracted galaxy densities by 
magnitude. The luminosity function of this sample is found 
to be steep, with a formal faint end slope of $\alpha = -2.2$ 
(Phillipps et al. 1998b), and as such is comparable with the 
steep functions found in some more distant clusters (e.g. 
Driver et al. 1994, De Propris et al. 1995, 
Smith, Driver \& Phillipps 1997, Wilson et al. 1997, Trentham 
1997, 1998a). At the bright end, galaxy numbers are consistent with 
those of Sandage, Binggeli and Tammann (1985). 
The dwarf density in the cluster core field is actually smaller 
than that in the outer field (430 galaxies deg$^{-2}$ against 
560~deg$^{-2}$). The dense environment in the cluster core, and 
particularly the presence of M87, may have a direct effect on the 
low surface brightness galaxy population, either through the 
removal of these galaxies or suppressing their formation. A similar 
effect has been found in the core of the Coma Cluster by 
Thompson \& Gregory (1993).

\section{The Full Cluster Survey}

    Progress is underway on reducing the full 100~deg$^{2}$ Virgo 
survey area. To date, 
23 of the required 24 exposures have been taken. 
It is intended that the full survey will measure the background 
population using data identical in format to those from Virgo, 
reducing systematic errors which might be introduced during the 
subtraction of the background galaxy numbers. 
Work is therefore progressing on the Virgo Northwest field 
and on a background field centred at right ascension 10h 40m, 
declination 0$^\circ 00'$ (1950). 
The photometric calibration has been accomplished using R band 
observations on the Anglo-Australian Telescope obtained for 
other projects.

For data reduction, each field is subdivided into 16 subregions, 
each $6800 \times 6800$ pixels. 
Data are stored as 4-byte numbers per pixel, and therefore 
each 1/16-th section is 185~Mbyte in size. 
New computer hardware resources provide 
sufficient memory to reduce each $6800 \times 6800$ pixel section 
as a whole, avoiding any need to break these sections into a large 
number of smaller regions as was done in the initial survey. 
This has the advantage of much simpler data handling than 
would be needed if large numbers of small sections were 
used. Similarly, establishing astrometric reference frames 
for a very large number of different sections would be 
prohibitive. 

As in the case of the initial survey of 3.2~deg$^2$, 
emphasis is being put on searches for small angular 
size low surface brightness dwarfs. However, 
the feasibility of performing a parallel survey with a different 
sky background 
subtraction is being investigated, in order to allow searches for 
large angular size low surface brightness galaxies of the type 
studied by Impey, Bothun and Malin (1988), but over a wider area 
than was available for their study.

\section{Conclusions}

     The new survey is already providing large samples of candidate 
Virgo Cluster dwarf spheroidal galaxies suitable for more detailed 
study. The galaxies have absolute magnitudes as faint as 
$M_{\rm R} = -11$ to $-16$ (for 
$H_0 = 70$~km$\:\mbox{s}^{-1}\:\mbox{Mpc}^{-1}$). 
They will be important for addressing 
questions relating to the properties and evolution of 
the lowest luminosity galaxies in an environment very different 
from that in nearby galaxy groups. 
An initial survey of 3.2~deg$^2$ has found a high density of 
these galaxies and evidence for a steep luminosity function. 
The survey is continuing over an area of 100~deg$^2$. 
The very large numbers of dwarf spheroidals should reduce the errors 
in the faint end of the luminosity function and help to define 
its shape. They will enable clear differences to be found at 
these very faint limits between the cluster core and periphery 
to significantly better accuracy than has been possible for any 
other cluster before now. A parallel survey of the Fornax Cluster 
is planned for which several films are already available. 

\acknowledgments
We wish to thank the UKST and Super{\small COSMOS} staff in 
providing their usual excellent service.

\end{document}